\documentclass[notitlepage,nofootinbib,preprintnumbers,aps,prd,superscriptaddress]{revtex4-1}
\usepackage{amsmath,amssymb,natbib,bm,color}
\usepackage{appendix}
\usepackage{graphicx}

\begin{document}
\preprint{\hbox{UTWI-19-2022}}
\title{Baryogenesis and Primordial Black Hole Dark Matter\\
from Heavy Metastable Particles}
\author{Barmak Shams Es Haghi}
\email{shams@austin.utexas.edu}
\affiliation{Texas Center for Cosmology and Astroparticle Physics, Weinberg Institute, Department of Physics, The University of Texas at Austin, Austin, TX 78712, USA}

\begin{abstract}
We propose a novel and simple scenario to explain baryon asymmetry and dark matter (DM) by utilizing an early matter-dominated era (EMDE) caused by a heavy metastable particle.  Within the EMDE, lack of pressure enhances the formation of primordial black holes (PBHs) which can then contribute to the relic abundance of DM. The eventual decay of heavy metastable particle that has baryon number and $CP$ violating interactions reheats the Universe and gives rise to baryon asymmetry. 
Since in this setup, PBH serves as a DM candidate, the particle physics model may not require new stable degrees of freedom which leads to more freedom in the model-building side. As an example, we show that a modulus field which dominates the energy density of the Universe prior to its decay, may explain both DM and baryon asymmetry in the Universe in the context of the Minimal Supersymmetric Standard Model (MSSM) while the lightest superparticle is not stable and cannot be a DM candidate due to the R-parity violating interactions needed for baryogenesis.
\end{abstract}
 
\maketitle
\section{Introduction}
The nature of dark matter (DM) and the origin of matter-antimatter asymmetry in the Universe constitute two of the major puzzles in both particle physics and cosmology.
There are myriad candidates for DM, ranging from ultra-light particles to heavy black holes, each one motivated by a certain production mechanism to lead to the observed relic abundance of DM today. 
To explain baryon asymmetry in the Universe, three necessary conditions should be met. These, the so-called Sakharov conditions~\cite{Sakharov:1967dj}, are: baryon number violation, $C$ and $CP$ violation, and departure from thermal equilibrium.
Although the electroweak sector of the Standard Model (SM) fulfills the first two conditions, the lack of a departure from thermal equilibrium in the SM excludes the possibility of explaining baryon asymmetry within the SM. Even if the electroweak phase transition, which is shown to be a smooth crossover, is assumed to be a first order phase transition, the predicted baryon asymmetry by the SM is too small. The failure of the SM to explain baryon asymmetry requires a dynamical mechanism of baryogenesis aided by beyond-SM physics. 

The intriguing observation that the abundances of DM and baryons are close to each other ($\Omega_\text{CDM}\sim 5 \Omega_B$), motivates the idea of a common origin or history for DM and baryon. Asymmetric DM \cite{Zurek:2013wia}, late-time decay of moduli~\cite{Allahverdi:2010rh,Kane:2011ih, Allahverdi:2013tca}, Affleck-Dine
scenario~\cite{Bell:2011tn, Borah:2022qln}, spontaneous matter genesis~\cite{March-Russell:2011ang,Kamada:2012ht}, splitting baryon number between quarks and antibaryons in a hidden sector via decay of a massive particle~\cite{Davoudiasl:2010am}, decay of a weakly interacting massive particle after its thermal freeze-out~\cite{Cui:2012jh}, Hawking evaporation of primordial black holes (PBHs) ~\cite{Fujita:2014hha}, formation of PBHs at QCD epoch~\cite{Carr:2019hud, Garcia-Bellido:2019vlf}, asymmetric DM collapsing into PBHs~\cite{Flores:2020drq}, baryogenesis triggered by inflation models for PBH formation~\cite{Wu:2021gtd}, and two population of light and heavy PBHs~\cite{Gehrman:2022imk} are some of these ideas. For a review of different baryogenesis scenarios and their possible connection to DM, see~\cite{Elor:2022hpa} and references therein.

In this paper, we show that the common link between DM and baryon asymmetry can be a heavy metastable particle that comes to dominate the energy density of the Universe prior to its decay and leads to an early matter-dominated era (EMDE). Within this EMDE, due to lack of pressure, the formation of PBHs is enhanced relative to the case of a radiation-dominated epoch. These PBHs can be heavy enough ($M_\text{PBH}\gtrsim 10^{15}\,\text{g}$) to survive to the present time and contribute to the final abundance of DM. The rest of the metastable particles which did not collapse into PBHs, eventually decay and reheat the Universe for the second time. 
Out of thermal equilibrium decay of the metastable particle can produce enough baryon asymmetry in the Universe provided that it has baryon number and $CP$ violating interactions with the SM~\cite{Weinberg:1979bt, Kolb:1979qa}. $CP$ violation necessitates complex couplings of the metastable particle to the SM. Since the complex phases are irrelevant at tree level, the interference between tree- and loop-level decays of the metastable particle is required to explain baryon asymmetry which makes the rate of this process loop suppressed. Hence, for Yukawa couplings of order one, the rate of baryon asymmetry production is expected to be reduced by a factor of $\mathcal{O}(10^{-2})$.

The idea of formation of PBHs within an EMDE was first introduced and studied in the context of grand unified theories~\cite{Khlopov:1980mg,1982SvA....26..391P,Polnarev:1985btg} as a cosmological trace of the new heavy particles predicted by these models. 
Recently this idea has been studied when perturbations seeding PBHs in the EMDE are sourced by inflaton field with a running spectral index or a spectator field which has a blue spectrum~\cite{Carr:2017edp}. Other studies of PBH formation in an EMDE can be found in Refs.~\cite{Kane:2015jia,Harada:2016mhb,Georg:2016yxa,Georg:2017mqk,Kokubu:2018fxy,Georg:2019jld,Matsubara:2019qzv}.
As long as we remain agnostic to the particle physics model, the mass of the heavy metastable particle, $m_X$, and its lifetime, $\tau_X$, are free parameters. The other free parameter of interest in this study is the fraction $\beta$ of the energy density of heavy metastable particles that collapses into PBHs.
These three parameters, in addition to the $CP$ violation required for baryogenesis, are subject to cosmological constraints including the age of the Universe, the observed abundance of DM and baryon asymmetry in the Universe, and the reheating temperature of the Universe\footnote{The beginning of the EMDE, which depends on the thermal history of the radiation-dominated era prior to the EMDE, can also affect the PBH mass function by setting the minimum possible mass of PBHs.}.
For PBHs to contribute to the abundance of DM today, they need to be heavy enough $(M_\text{PBH} \gtrsim 10^{15}\,\text{g}$) to have a lifetime longer than or of the order of the age of the Universe.
A successful reheating requires a reheating temperature above MeV scale~\cite{Kawasaki:1999na,Kawasaki:2000en,Hannestad:2004px,Ichikawa:2005vw,Ichikawa:2006vm,deSalas:2015glj,Hasegawa:2019jsa} to be able to drive Big Bang Nucleosynthesis (BBN) successfully. 
In the absence of $B-L$ violating interactions, any baryon asymmetry produced before Electroweak phase transition will be washed out by sphaleron. While by lowering the reheating temperature of the Universe below the electroweak scale, one can avoid the washout, adding $B-L$ violating interactions can relax this constraint.

As we show, the abundance of PBHs is independent of their mass and only depends on their initial abundance at formation time and the lifetime of the metastable particle. The generated baryon asymmetry depends on the mass of the heavy metastable particle, its lifetime, and the $CP$ violating parameter which encapsulates the amount of baryon asymmetry production and is model-dependent. Since the abundances of DM and baryon depend on the lifetime of the metastable particle similarly, fixing particle physics to explain baryon asymmetry also fixes the required initial abundance of PBHs to explain DM today. In other words, for a specific $CP$ violating parameter, and for a given choice of any one of the three parameters $m_X, \tau_X, \beta$, the other two are also determined in order to explain the DM and baryon asymmetry.
Due to these constraints, the parameter space that works is restricted.
 
We find that in order for PBHs to survive to the present time, the lifetime of the heavy metastable particle needs to be at least $\simeq 10^5\,\text{GeV}^{-1} (6.6\times10^{-20}\,\text{s})$. The mass of PBHs form before this time can be at most equal to $\simeq 10^{15}\,\text{g}$ where with an initial abundance equal to $\simeq 2.7\times 10^{-16}$, they are able to explain all the DM today. To produce baryon asymmetry (for order one Yukawa couplings involved in the decay), this particle needs to have a mass of $\simeq 1.9\times 10^{14}\,\text{GeV}$. Furthermore, to avoid sphaleron washout, the lifetime of the particle has to be larger than $\simeq 4.7\times 10^{13}\,\text{GeV}^{-1} (3.1\times 10^{-11}\,\text{s})$ (This constraint is not relevant in the presence of $B-L$ violating interactions). This corresponds to a maximum possible mass of $\simeq 9.6\times 10^{24}\,\text{g}$ for PBHs where they can explain all DM today for initial abundance of $5.8\times 10^{-12}$. Baryon asymmetry from the decay of this particle would be sufficient if it has a mass of $\simeq 8.6\times 10^9\,\text{GeV}$. And finally, requiring a reheating temperature to be above MeV scale demanded by BBN, leads to an upper limit on the lifetime of the particle which is equal to $\simeq 4.7\times 10^{23}\,\text{GeV}^{-1} (0.3\,\text{s})$. PBHs formed when a particle with this lifetime dominates the Universe can have a maximum mass of $\simeq 3\times 10^{36}\,\text{g}$ and with an initial abundance of $\simeq 5.8\times 10^{-7}$, they are capable of explaining the DM today. This particle can also generate enough baryon asymmetry after its decay when it has a mass of $\simeq 86\,\text{TeV}$. In this paper, we focus on the possible mass range for PBHs, but we note that the PBH extended mass function depends on the power spectrum of primordial density fluctuations which can possibly be tuned to explain all or most of the DM abundance while it is still consistent with various constraints on heavy PBHs.

As a realization of this scenario, we consider a modular cosmology, motivated by string theory and supergravity, within which a modulus field dominates the Universe before its decay. PBHs may form during this modulus-dominated era and be the DM candidate. The decay chain of the modulus field, consists of modulus decaying into gluino and then gluino decaying through R-parity violating operators, provides necessary baryon asymmetry in the Universe. A modulus field in the $100-10^4\,\text{TeV}$ mass range can produce enough baryon asymmetry and can lead to formation of PBHs where the upper bound on their mass lies in the range $10^{29}-10^{36}\,\text{g}$.
Due to the R-parity violating operators, the lightest supersymmetric particle cannot be a viable DM candidate, although this issue has already been addressed: PBHs can obviate the need for a particle DM candidate~\cite{Gehrman:2022imk}. 

The outline of this paper is as follows. In Section~\ref{sec:energy}, we describe our scenario, which includes an EMDE caused by a heavy metastable particle and its transition into a radiation-dominated era. In Section~\ref{sec:PBH}, after reviewing the formation of PBHs in an EMDE and their possible mass range, we evaluate the abundance of them today. In Section~\ref{sec:baryogenesis}, we obtain the baryon asymmetry in the Universe generated through decay of the heavy metastable particle. In Section~\ref{sec:constraints} we present and discuss the relevant cosmological constraints. Our results are discussed in Section~\ref{sec:resultss},
and finally in Section~\ref{sec:moduli}, we use a modulus field, as an illustration of our scenario.

\section{Evolution of Energy Content of the Universe}
\label{sec:energy}
A heavy metastable particle, $X$, comes to dominate the Universe at $t=t_{X\text{D}}$ and triggers an EMDE. We will  take the total energy density of the Universe at this time to consist mostly of $X$ particles, i.e. we assume the radiation component to be negligible, so that
\begin{equation}
    \rho_\text{tot}(t_{X\text{D}})\simeq \rho_X(t_{X\text{D}}).
\end{equation}
Within this EMDE, and before the decay of $X$ particles, PBHs can form at some time $t=t_\text{PBH}$. 
Just before formation of PBHs, at $t=t_\text{PBH}-\epsilon$, the energy density of the Universe which has been diluted by the expansion of the Universe, is still composed of $X$ particles:
\begin{equation}
    \rho_\text{tot}(t_\text{PBH}-\epsilon)=\rho_X(t_\text{PBH}-\epsilon),
\end{equation}
where $t\mp\epsilon$ is used to mark ``just before" and ``immediately after" some certain moment, $t$.

Immediately after formation of PBHs, at $t=t_\text{PBH}+\epsilon$, a fraction $\beta$ of the energy density of the Universe collapses into PBHs:
\begin{equation}
    \rho_\text{tot}(t_\text{PBH}+\epsilon)=\rho_\text{PBH}(t_\text{PBH}+\epsilon)+\rho_X(t_\text{PBH}+\epsilon)=\beta\rho_X(t_\text{PBH}-\epsilon)+(1-\beta)\rho_X(t_\text{PBH}-\epsilon).
\end{equation}
Since the energy density of PBHs is diluted by Universe expansion similar to the energy density of non-relativistic particles, the cosmology does not change and the EMDE continues until the $X$ particles eventually decay at $t=t_\text{dec}$ and reheat the Universe.

Right before decay of $X$ at $t=t_\text{dec}-\epsilon$, total energy density of the Universe includes $X$ particles and PBHs:
\begin{equation}
    \rho_\text{tot}(t_\text{dec}-\epsilon)=\rho_X(t_\text{dec}-\epsilon)+\rho_\text{PBH}(t_\text{PBH}-\epsilon).
\end{equation}
At $t=t_\text{dec}+\epsilon$, $X$ particles decay into radiation (lighter particles, including the SM particles). The energy content of the Universe at this moment includes radiation and PBHs:
\begin{equation}
    \rho_\text{tot}(t_\text{dec}+\epsilon)=\rho_\text{rad}(t_\text{dec}+\epsilon)+\rho_\text{PBH}(t_\text{PBH}+\epsilon).
\end{equation}
Radiation reheats the Universe to a temperature $T_\text{rh}$ which by using conservation of energy and also assuming instantaneous thermalization of decay products, can be evaluated by:
\begin{equation}
    \rho_\text{rad}(t_\text{dec}+\epsilon)=\rho_X(t_\text{dec}-\epsilon)\equiv \frac{\pi^2}{30}g_\star(T_\text{rh})T_\text{rh}^4,
\end{equation}
where $g_\star(T)$ counts the total number of relativistic degrees of freedom at temperature $T$.

During the time interval $t_{X\text{D}}\lesssim t \lesssim t_\text{dec}$, the Universe is filled with matter ($X$ particles, or $X$ particles plus PBHs), so at $t=t_\text{dec}\simeq \tau_X$ the Hubble expansion rate, $H$ is given by
\begin{equation}
    H(t_\text{dec})=\frac{2}{3 t_\text{dec}}\simeq \frac{2}{3\tau_X}.
\end{equation}
By using the Friedmann equation, $H^2 =8\pi\rho_\text{tot}/(3M_\text{Pl}^2)$, one can obtain the energy density of the Universe at $t_\text{dec}$ as:
\begin{equation}
    \rho_\text{tot}(t_\text{dec})=\frac{1}{6\pi}\left(\frac{M_\text{Pl}}{\tau_X}\right)^2,
\end{equation}
Since the energy density of $X$ and PBHs, $\rho_X$ and $\rho_\text{PBH}$ respectively, are diluted with the same rate by Universe expansion ($\sim a^{-3}$), at $t_\text{dec}$ the same fraction of energy density turns into radiation that did not collapse into PBHs at $t_\text{PBH}$, i.e., $(1-\beta)$. Therefore
\begin{equation}
  \rho_\text{rad}(t_\text{dec})=(1-\beta) \rho_\text{tot}(t_\text{dec})= \frac{1-\beta}{6\pi}\left(\frac{M_\text{Pl}}{\tau_X}\right)^2,
\end{equation}
and equivalently, the reheating temperature, $T_\text{rh}$ is given by:
 \begin{equation}
     T_\text{rh}=\frac{5^{1/4}}{\pi^{3/4}}\frac{(1-\beta)^{1/4}}{g_\star(T_\text{rh})^{1/4}}\sqrt{\frac{M_\text{Pl}}{\tau_X}}.
     \label{eq:Trh}
 \end{equation}

\section{PBH formation and Relic Abundance}
\label{sec:PBH}
During an EMDE, absence of pressure enhances PBH formation. The Jeans length is much
smaller than the particle horizon, and instead of pressure, the main hurdles for PBH formation are inhomogeneity and anisotropy (deviations from spherical symmetry)~\cite{Khlopov:1980mg}. More specifically, almost spherical overdensities can collapse into PBHs while non-spherical collapse of fluctuations leads to formation of two-dimensional pancakes and their virialisation stops them from turning into PBHs~\cite{Harada:2016mhb, Kokubu:2018fxy}.
By including the effects of these two factors, the probability of formation of PBHs during an EMDE is shown to be equal to~\cite{Khlopov:1980mg}:
\begin{equation}
    \beta(M_\text{PBH}) \approx 0.02\sigma^{13/2}(M_\text{PBH}),
    \label{eq:beta1}
\end{equation}
where $\sigma(M_\text{PBH})$ is the density
fluctuation at horizon entry. The more recent analysis which argues that inhomogeneity effect is model-dependent, by using  Zel'dovich approximation and only including non-spherical effect, confirms this result
~\cite{Harada:2016mhb}:
\begin{equation}
    \beta(M_\text{PBH}) \approx 0.05556\sigma^5(M_\text{PBH}),
    \label{eq:beta2}
\end{equation}
where the extra factor of $\sigma^{3/2}$ in Eq.~(\ref{eq:beta1}) counts for the probability for sufficient homogeneity~\cite{Khlopov:1980mg}.  An initial fluctuation can give rise to the formation of a PBH if the deviation of spherical symmetry is smaller than its own size. This implies five conditions for independent components of tensor of deformation before its diagonalization: three diagonal components need to be close to each other (two conditions) and non-diagonal components should be be small (three conditions). Roughly speaking, this is the reason that deviation from spherical symmetry leads to $\beta\sim\sigma^5$~\cite{Khlopov:2008qy}.
The formation probability of PBHs with spins can be found in Ref.~\cite{Harada:2017fjm}. 
The evolution of matter in the absence of pressure within a matter-dominated era, which follows collisionless Boltzmann equation, may develop velocity dispersion in the nonlinear regime. The effect of velocity dispersion on PBH formation in a matter-dominated era has been studied recently in Ref~\cite{Harada:2022xjp}.

Depending on the time of formation of PBHs which can be any time within the range $t_{X\text{D}}\lesssim t \lesssim \tau_X$, the mass of the PBHs follows the horizon mass, $M_\text{H}$. It is shown~\cite{Khlopov:1980mg} that PBH formation enhances over the following mass range:
\begin{equation}
    M_\text{min}\sim M_\text{H}(t_{X\text{D}})\lesssim  M_\text{PBH}\lesssim M_\text{max}\sim M_\text{H}(\tau_X)\sigma^{3/2}(M_\text{max}),
    \label{eq:massrange}
\end{equation}
where $M_\text{max}$ is determined by requiring that fluctuations in the considered scale $M_\text{max}$, enter the horizon at $t(M_\text{max})$ such that they grow to nonlinear regime and decouple before $\tau_X$, i.e., \cite{Khlopov:1980mg, Khlopov:2008qy}:
\begin{equation}
    \sigma(M_\text{max})=[t(M_\text{max})/\tau_X]^{2/3}.
\end{equation}

During the time interval, $t_{X\text{D}}\lesssim t \lesssim \tau_X$, the mass of the formed PBHs in the EMDE caused by $X$ particle, falls into the following range:
\begin{equation}
 \frac{3}{4} M^2_\text{Pl} t_{X\text{D}}\lesssim M_\text{PBH}\lesssim \frac{3}{4} M^2_\text{Pl}  \tau_X \sigma^{3/2}(M_\text{max}).
 \label{eq:massrange}
\end{equation}
PBHs eventually evaporate due to Hawking evaporation. If their lifetime is at least equal to the age of the Universe, they can contribute to the abundance of DM today. We require all the PBHs formed during the EMDE to be heavy enough, i.e., $M_\text{PBH}\gtrsim 10^{15}\text{g}$, to survive until today, or from Eq.~(\ref{eq:massrange}), $t_{X\text{D}}\gtrsim 5\,\text{GeV}^{-1}\simeq 3.3\times 10^{-24}\,\text{s}$.
For consistency, we also need $t_{X\text{D}}\lesssim \tau_X \sigma^{3/2}(M_\text{max})$, otherwise perturbations do not have enough time to grow during the EMDE and therefore no PBH can form.
The relic abundance of PBHs today, can be evaluated as: 
 \begin{eqnarray}
   \nonumber \Omega_\text{PBH}&=&\frac{\rho_\text{PBH}(t_0)}{\rho_c(t_0)}=\left[\frac{a(\tau_X)}{a(t_0)}\right]^3\frac{\rho_\text{PBH}(\tau_X)}{\rho_c(t_0)}=\frac{s(t_0)}{s(\tau_X)}\frac{\rho_\text{PBH}(\tau_X)}{\rho_c(t_0)}= \frac{3}{4}\frac{g_{\star}(T_\text{rh})}{g_{\star,S}(T_\text{rh})}\frac{\rho_\text{PBH}(\tau_X)}{\rho_\text{rad}(\tau_X)}\frac{s(t_0)}{\rho_c(t_0)}T_\text{rh}\\
   &=&\frac{3}{4}\frac{g_{\star}(T_\text{rh})}{g_{\star,S}(T_\text{rh})}\frac{\beta}{1-\beta}\frac{s(t_0)}{\rho_c(t_0)}T_\text{rh}=\frac{3\times 5^{1/4}}{4\pi^{3/4}}
     \frac{g_{\star}(T_\text{rh})^{3/4}}{g_{\star,S}(T_\text{rh})}
     \frac{\beta}{(1-\beta)^{3/4}}\sqrt{\frac{M_\text{Pl}}{\tau_X}}
     \frac{s(t_0)}{\rho_c(t_0)},
     \label{eq:OmegaPBH}
 \end{eqnarray}
 where $s(t)$ is the entropy density of the Universe, $\rho_c$ is the critical energy density of the Universe, $g_{\star,S}(T)$ is the number of relativistic degrees of freedom contributing to the entropy of the Universe, and $t_0$ denotes the present time.  We have 
 $\rho_c(t_0)=1.0537\times 10^{-5}\, h^2\,\,\rm{GeV}~{\rm cm}^{-3}$ and $s(t_0)=2891.2\left({T_0}/{2.7255 {\rm K}}\right)^3 \rm{cm^{-3}}$~\cite{Planck:2018vyg} where $h=0.674$ is scaling factor for Hubble expansion rate.
 We note the dependence on our key parameters, 
 \begin{equation}
 \Omega_\text{PBH} \propto \frac{\beta}{\sqrt{\tau_X}},
 \end{equation}
 for $\beta \ll 1$ as we shall require below to explain DM abundance.

 \section{Baryogenesis}
 \label{sec:baryogenesis}
 It is customary to quantify the baryon asymmetry introduced by $X$ particles with the $CP$ violating parameter, $\gamma_{CP}$, which is defined as:
\begin{equation}
   \gamma_{CP}=\sum_i B_i\frac{\Gamma(X\rightarrow f_i)-\Gamma(\bar{X}\rightarrow \bar{f_i})}{\Gamma_X},
\end{equation}
where $B_i$ is the baryon number of the particular final state $f_i$, and $\Gamma_X$ is the $X$ particle decay width. The magnitude of baryon asymmetry is model-dependent, but it is reasonable to assume $\gamma_{CP}\sim 10^{-2}$ (due to loop suppression) without being specific about the details of the beyond-SM physics.

The baryon-number-to-entropy density is given by:
\begin{eqnarray}
     \nonumber Y_B &=&\frac{n_B(t_0)}{s(t_0)}=\frac{n_B(t_\text{dec}+\epsilon)}{s(t_\text{dec}+\epsilon)}=\gamma_{CP}\frac{n_X(t_\text{dec}-\epsilon)}{s(t_\text{dec}+\epsilon)}=
     \frac{3}{4}\gamma_{CP}\frac{g_{\star}(T_\text{rh})}{g_{\star,S}(T_\text{rh})}\frac{T_\text{rh}}{m_X}
     \frac{\rho_X(t_\text{dec}-\epsilon)}{\rho_\text{rad}(t_\text{dec}+\epsilon)}\\
     &=&\frac{3}{4}\gamma_{CP}\frac{g_{\star}(T_\text{rh})}{g_{\star,S}(T_\text{rh})}\frac{T_\text{rh}}{m_X}
     \frac{\rho_X(t_\text{dec}-\epsilon)}{\rho_X(t_\text{dec}-\epsilon)},
\end{eqnarray}
which leads to
\begin{equation}
    Y_B=\frac{3}{4}\gamma_{CP}\frac{g_{\star}(T_\text{rh})}{g_{\star,S}(T_\text{rh})}\frac{T_\text{rh}}{m_X}
    =\frac{3\times 5^{1/4}}{4\pi^{3/4}}
     \frac{g_{\star}(T_\text{rh})^{3/4}}{g_{\star,S}(T_\text{rh})}\gamma_{CP}
     (1-\beta)^{1/4}\frac{1}{m_X}\sqrt{\frac{M_\text{Pl}}{\tau_X}}.
     \label{eq:YB}
\end{equation}
By using the parameter $Y_B$ and the mass of the proton, $m_p$, the abundance of baryons today, $\Omega_B$, is obtained as:
\begin{equation}
    \Omega_B=\frac{\rho_B(t_0)}{\rho_c(t_0)}=\frac{m_p n_B(t_0)}{\rho_c(t_0)}=\frac{m_ps(t_0)}{\rho_c(t_0)}Y_B.
    \label{eq:OmegaB}
\end{equation}
We note the dependence on our key parameters,
\begin{equation}
\Omega_B \propto Y_B \propto \frac{\gamma_{CP}}{m_X\sqrt{\tau_X}},
\end{equation}
for the case $\beta \ll 1$.

From Eqs.~(\ref{eq:OmegaPBH}) and~(\ref{eq:OmegaB}), one can find out the ratio of the abundance of PBHs to the abundance of baryon as:
\begin{equation}
    \frac{\Omega_\text{PBH}}{\Omega_B}=\frac{1}{\gamma_{CP}}\frac{\beta}{1-\beta}\frac{m_X}{m_p}\simeq\frac{\beta}{\gamma_{CP}}\frac{m_X}{m_p}.
\end{equation}
Since $\Omega_\text{PBH}$ and $\Omega_B$ have the same dependence on $\tau_X$ (see Eqs.~(18) and (23)), their ratio is independent of the lifetime of $X$. 

It is worth mentioning that $\beta$ required to explain DM today, is uniquely determined by the underlying particle physics ($m_X,\tau_X, \gamma_{CP}$) which is responsible for baryogenesis. For a fixed $\gamma_{CP}$, any one of the three parameters $m_X, \tau_X, \beta$, determines the value of the other two in order to explain observed abundances of the DM and baryon.
As we will see the parameter space that works is very restricted by the constraints.

\section{Cosmological Constraints}
\label{sec:constraints}
In this section, we present the relevant cosmological constraints on our model, namely the lifetime and relic abundance of the DM today, the observed amount of baryon asymmetry in the Universe, the requirements of BBN, and the capture of heavy metastable particle by PBHs before decay.  Subsequently, in the next section Section~\ref{sec:resultss}, we will apply these constraints to our model, to find bounds on the free parameters in this study (mass and lifetime of $X$, $m_X$ and $\tau_X$ respectively, and the initial abundance of PBHs, $\beta$). 
\subsection{Dark Matter: Stability and Abundance}
PBHs can contribute to the final abundance of DM if their lifetime is larger than the age of the Universe; otherwise they would disappear due to Hawking evaporation. For Schwarzschild PBHs, this leads to a lower bound on their mass: 
\begin{equation}
 M_\text{PBH} \gtrsim 10^{15}\,\text{g}.
\end{equation}
The above bound varies, but not substantially, by including the spin of PBHs~\cite{Arbey:2019jmj}.

To avoid overclosing the Universe, the abundance of PBHs needs to be smaller than or equal to the observed abundance of cold DM~\cite{Planck:2018vyg}:
\begin{equation}
\Omega_\text{PBH} h^2 \leq \Omega_\text{CDM} h^2 = 0.12.
\end{equation}
\subsection{Baryogenesis}
The observational value of baryon-number-to-entropy density, $Y_{B,\,\text{obs}}$, is~\cite{Zyla:2020zbs}:
\begin{equation}
   Y_{B,\,\text{obs}}\simeq 8.7\times 10^{-11}.
   \label{eq:YBobs}
\end{equation}
For temperatures above the electroweak phase transition, $T\geq T_\text{EW}\sim100\,\text{GeV}$, up to $T\sim 10^{12}\,\text{GeV}$~\cite{Khlebnikov:1988sr}, sphaleron processes are in thermal equilibrium and can erase any baryon asymmetry. To avoid the washout due to sphaleron, we require $T_\text{rh}<T_\text{EW}$. This constrain can be relaxed by introducing $B-L$ violating interactions.

A heavy metastable particle with large enough energy density prior to its decay, or equivalently large enough decay width, can lead to a reheating temperature larger than its own mass. To evade the consequent washout, we need $T_\text{rh}<m_X$. The observed baryon asymmetry in the Universe given by Eq.~(\ref{eq:YBobs}), together with Eq.~(\ref{eq:YB}), show that this condition is always satisfied:
\begin{equation}
    \frac{T_\text{rh}}{m_X}\simeq 10^{-8}\left(\frac{10^{-2}}{\gamma_{CP}}\right)\frac{Y_B}{Y_{B,\,\text{obs}}}.
    \end{equation}

\subsection{Big Bang Nucleosynthesis}

For the BBN to proceed successfully, the reheating temperature of the Universe must be higher than the MeV scale.
This constraint, through Eq.~(\ref{eq:Trh}), sets an upper bound on $\tau_X$.
\subsection{Capture of Heavy Metastable Particles by PBHs}
PBHs formed within the EMDE caused by $X$ particles might capture some of them before these particles decay and reheat the Universe.
To find out when this capture process becomes important, we evaluate the ratio of the number of captured particles, $N_{X_\text{Capt}}$, during a Hubble time, $t_H$, to the total number of $X$ particles, $N_X$, within a Hubble volume, $V_H$~\cite{Stojkovic:2004hz, Gondolo:2020uqv},  
\begin{equation}
\frac{N_{X-\text{Capt}}}{N_X}=\frac{(n_X\sigma_{X,\text{PBH}} v_X t_H) n_\text{PBH} V_H}{n_X V_H}=\frac{n_\text{PBH}\sigma_{X,\text{PBH}} v_X}{H}.
\label{eq:captrate}
\end{equation}
Here $n_X$ is the number density of $X$ particles, $n_\text{PBH}$ is the number density of PBHs, $v_X$ is the velocity of an $X$ particle, $H$ is the Hubble rate, and $\sigma_{X,\text{PBH}}$ is the cross-section for gravitational capture of a non-relativistic particle by a black hole. For a non-rotating black hole, $a_\star=0$, where $a_\star=J M_\text{Pl}^2/M_\text{BH}^2$ is the dimensionless black hole
angular momentum and $J$ is the angular momentum of the black hole; in this case the gravitational capture cross-section is given by 
\begin{equation}
\sigma_{X,\text{PBH}}=\frac{4\pi r_\text{S}^2}{v_X^2}=\frac{16\pi}{v_X^2}\frac{M_\text{PBH}^2}{M_\text{Pl}^4},
\label{eq:crosssec}
\end{equation}
where $r_\text{S}$ is the Schwarzschild radius of the black hole.
PBHs form within an EMDE tend to have large spins~\cite{Harada:2017fjm}. The cross-section for gravitational capture of non-relativistic particles by an extremely rotating black hole, $a_\star=1$, when particles fall perpendicularly and parallel to the rotation axis of the black hole are $14.2/16$ and $14.8/16$ (respectively) times the capture cross-section of non-rotating black hole with the same mass, i.e., Eq.~(\ref{eq:crosssec}); non-rotating black holes capture particles with higher efficiency than rotating black holes with the same mass~\cite{Frolov:1998wf}.
Here, to estimate the maximum capture fraction, we use the upper bound on the capture cross-section which corresponds to the non-rotating black holes, we also assume a monochromatic mass function for PBHs peaked at the maximum possible mass for the PBHs. Following Eq.~(\ref{eq:massrange}), we obtain:
\begin{equation}
\frac{N_{X-\text{Capt}}}{N_X}=\frac{3\beta\sigma^{3/2}}{v_X}\simeq7.1\frac{\beta^{13/10}}{v_X},
\end{equation}
which shows that when heavy particles move very slowly or for larger values of the initial abundance of PBHs, the capture rate can be noticeable and needs to be included in the analysis.

\section{Results}
\label{sec:resultss}
\begin{figure}[t]
  \centering
    \includegraphics[width=0.5\textwidth]{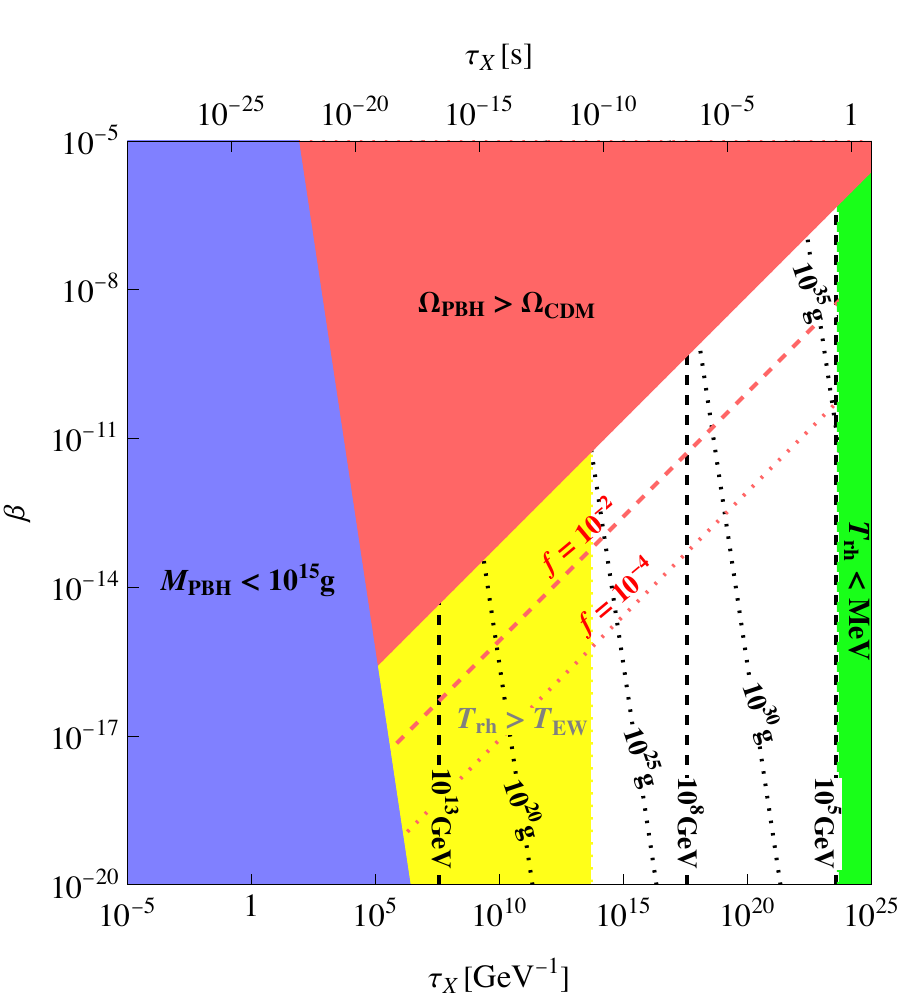}
  \caption{Constraints on our model in the $(\tau_X, \beta)$ plane, where $\tau_X$ is the lifetime of the metastable particle (responsible for the EMDE) and $\beta$ is the initial fraction of the energy density of the Universe that collapses into PBHs. Dotted black contours display the upper bound on the mass of PBHs that can form when $X$ particles dominate the Universe. The blue, red, and green shaded regions are excluded by the lifetime of PBHs as a viable DM candidate, the abundance of PBHs today, and BBN respectively. The yellow shaded region corresponds to $T_\text{rh}>T_{EW}$ where the sphaleron processes are in thermal equilibrium and can wash out any baryon asymmetry. Presence of $B-L$ violating interactions opens up this excluded region. Vertical dashed black contours show some benchmark values of $m_X$ which can explain baryon asymmetry in the Universe for $\gamma_{CP}=10^{-2}$ and for some certain value of $\tau_X$.
  While Formation of PBHs with a mass larger than $10^{15}\,\text{g}$ is enhanced within the white region (and the yellow region with $B-L$ violating interactions), they can account for a fraction $f\equiv\Omega_\text{PBH}/\Omega_\text{CDM}<1$ of DM relic abundance. Dashed and dotted red contours depict $f=10^{-2}$ and $f=10^{-4}$ respectively. In order for the PBHs to explain the full DM abundance of our Universe, the parameters must lie along the inclined right-hand edge of the red region.}
  \label{fig:parameterspace}
\end{figure}

Fig.~\ref{fig:parameterspace} shows our primary results including the excluded regions in the $(\tau_X,\beta)$ parameter space by different cosmological constraints as described in Section~\ref{sec:constraints}. 

The dotted black contours in Fig.~\ref{fig:parameterspace} show the maximum possible mass of PBHs that can form within an EMDE caused by $X$ particles of lifetime $\tau_X$. The maximum possible mass is obtained from the upper bound in Eq.~(\ref{eq:massrange}) and also Eq.~(\ref{eq:beta2}), assuming most of PBHs have the maximum mass. In reality, PBHs have an extended mass function with a high mass cut-off which is less than the reported numbers here since PBHs with lower masses also contribute to the abundance of PBHs. In the blue shaded region, the maximum mass of PBHs is less than $10^{15}\,\text{g}$, their lifetime is shorter than the  age of the Universe, and therefore they cannot contribute to the relic abundance of DM today. The red shaded region is excluded to avoid overclosing the Universe by PBHs; below the red shaded region PBHs are underproduced and only constitute a fraction $f\equiv\Omega_\text{PBH}/\Omega_\text{CDM}<1$ of DM today. For example along the dashed and dotted red lines, $f=10^{-2}$ and $f=10^{-4}$ respectively. 
In order for the PBHs to explain the
full DM abundance, the parameters must lie along the inclined right-hand edge of the red region.
The lifetime of PBHs and their relic abundance together require that the lifetime of the heavy metastable particle to be at least $\simeq 10^5\,\text{GeV}^{-1} (6.6\times10^{-20}\,\text{s})$.

Within the yellow shaded region, the reheating temperature of the Universe due to decay of the $X$ particles is above the electroweak phase transition and sphaleron processes can erase any baryon asymmetry. While evading sphaleron washout leads to a lower bound of $\simeq 4.7\times 10^{13}\,\text{GeV}^{-1} (3.1\times 10^{-11}\,\text{s})$ on the lifetime of the particle, this constraint can be totally bypassed by adding $B-L$ violating interactions. Finally the green shaded region is excluded by BBN; demanding a reheating temperature above MeV scale results in an upper limit on the lifetime of the particle which is equal to $\simeq 4.7\times 10^{23}\,\text{GeV}^{-1} (0.3\,\text{s})$.

The vertical dashed black contours in Fig.~\ref{fig:parameterspace} represent some benchmark values of $m_X$ that can explain baryon asymmetry in the Universe for $\gamma_{CP}=10^{-2}$; for a fixed particle physics model, i.e., for a fixed $\gamma_{CP}$, the value of $\tau_X$ determines uniquely the value of $m_X$ and $\beta$ that can explain baryon asymmetry and the relic abundance of DM today, respectively. $\beta$ together with $\tau_X$ approximate the maximum mass of PBHs.
For instance, for $m_X=10^{8}\,\text{GeV}$, a lifetime of $\tau_X\simeq 3.5\times 10^{17}\,\text{GeV}^{-1} (2.3\times10^{-7}\,\text{s})$ is needed to produce the observed baryon asymmetry in the Universe. For this lifetime, PBHs with initial abundance of $\beta\simeq 5\times 10^{-10}$ can explain the DM abundance today and their mass can be as large as $\simeq 2.7\times10^{29}\,\text{g}$. It is worth noting that to explain DM relic density, $\beta$ is independent of the mass of PBHs, while the upper limit on the mass of PBHs depends on $\beta$. 

We notice that heavy PBH as DM candidate is subject to various astrophysical and cosmological constraints~\cite{Carr:2016drx,Carr:2020gox,Carr:2020xqk,Green:2020jor,Villanueva-Domingo:2021spv, Auffinger:2022khh, Escriva:2022duf}. These constraints are evaluated for monochromatic PBH mass function. Since PBHs usually form with an extended mass function rather than a monochromatic one, the constraints need to be re-evaluated for the extended mass function of interest~\cite{Carr:2016drx, Kuhnel:2017pwq, Bellomo:2017zsr}. While it is difficult to explain all the DM abundance by PBHs with monochromatic mass function, it may be still possible if the mass function is extended~\cite{Carr:2016drx, Kohri:2018qtx}. Within an EMDE, the finite duration of the matter dominance leads to cut-offs on the mass of PBHs. The details of the the power spectrum of density fluctuations which are responsible for PBH formation determines the shape of the PBH mass function. For instance, for scale-invariant primordial fluctuations, $\beta$ is almost constant, and the duration of EMDE can affect how wide or narrow the PBH mass function might be~\cite{Carr:2018rid, Carr:2018nkm}, or the PBH mass function would skew toward the minimum (maximum) PBH mass cut-off for a blue(red)-tilted power spectrum~\cite{ Carr:2018nkm, Kohri:2018qtx}. As it is shown in Ref.~\cite{Kohri:2018qtx}, for inflation with large running spectral indices, parameters can be tuned such that PBHs form within an EMDE would be able to explain all the DM abundance with an extended mass function which is consistent with the constraints; tuning the parameters for reheating temperatures up to $T_\text{rh}\simeq 10^4\,\text{GeV} (\tau_X\gtrsim 10^9\,\text{GeV}^{-1})$ can result in PBH mass functions which are allowed by constrains and can explain all the DM abundance. We emphasize that in this study, PBH formation is demanded to be enhanced over $10^{15}\,\text{g}\lesssim M_\text{PBH}\lesssim M_\text{PBH, max}(\tau_X,\beta)$ mass range and fixing free parameters of our study does not determine the PBH mass function. The exact shape of PBH mass function also depends on the onset of matter dominance, $t_{X\text{D}}$, and 
the power spectrum of primordial density fluctuations. Although explaining DM just by heavy PBHs is challenging, these extra parameters
can potentially be tuned~\cite{Kohri:2018qtx}, to make the mass function peak within the mass window that is not (less) constrained (see Ref.~\cite{Escriva:2022duf}) and explain all (most of) the relic abundance of DM today. Since PBH mass function depends on the choice of the power spectrum, we do not include astrophysical and cosmological constraints here and 
leave a more careful study to future work.
\section{Particle Physics Model: Modulus Field}
\label{sec:moduli}

An example of an EMDE caused by a heavy metastable particle is when the Universe is dominated by the oscillations of moduli fields, which are ubiquitous in supergravity and string theory, and have only gravitational interactions~\cite{Coughlan:1983ci}. 
In general, modulus field may be displaced from the minimum of its potential and starts to oscillate. The energy density of modulus field is diluted like energy density of matter by the expansion of the Universe. It eventually dominates the Universe and causes a transition from a radiation-dominated era to a matter-dominated era. During this modulus-dominated epoch, heavy or light PBHs can form. For studies of formation of light PBHs and solar mass PBHs in a modulus-dominated epoch, see~\cite{Arbey:2021ysg} and~\cite{Bhattacharya:2021wnk} respectively. 
The weakening effects of formation of PBHs in an EMDE caused by moduli on the constraints on density perturbations can be found in Ref.~\cite{Green:1997pr}.

Modulus field eventually decays and reheats the Universe for a second time. The decay chain of a modulus field can successfully explain the baryon asymmetry of the Universe~\cite{Kitano:2008tk, Allahverdi:2010im, Allahverdi:2010rh, Ishiwata:2013waa, Dhuria:2015xua,Kane:2019nes}. 

The total decay width
of the modulus field of mass $m_X$ is given by:
\begin{equation}
    \Gamma_X\simeq c \frac{m_X^3}{M_\text{Pl}^2},
\end{equation}
where $c\sim \mathcal{O}(1)$.
The possible sizable branching ratio of decay of modulus into gluinos, $\Tilde{g}$, paves the way for producing baryons abundantly via gluino decay into quark and squark. 
A successful baryon asymmetry production at the end of a modular cosmology can occur by adding R-parity violating renormalizable operators to the superpotential of the Minimal Supersymmetric Standard Model (MSSM) such as~\cite{Ishiwata:2013waa, Kane:2019nes}
\begin{equation}
  W \supset\lambda^{''}_{ijk}\epsilon _{lmn}U^{cl}_i D^{cm}_j D^{cn}_k,
\end{equation}
where $U^c$ and $D^c$ denote the $SU(2)_L$ singlet up-type and down-type quark superfields, respectively, $i, j, k = (1, 2, 3)$ represent flavor indices while $l, m, n$ are color indices, and $\epsilon _{lmn}$ is the Levi-Civita tensor. 

Following~\cite{Kane:2019nes}, we assume $\lambda^{''}_{ijk}=0$ except for $\lambda^{''}_{323}=-\lambda^{''}_{332}$, which can be large, and therefore the baryon number violation occurs through $\Tilde{g}\rightarrow t+ s +b \,( \bar t+\bar s +\bar b)$. $CP$ violation is originated from phase difference between gluino and bino, $\tilde{B}$ (or wino, $\tilde{W}$) or phases in R-parity violating couplings.
We further assume that all sfermions other than the lightest stop, $\Tilde{t}_1$, are decoupled. The $CP$ violating parameter of gluino, $\gamma_{CP,\Tilde{g}}$, for a purely right-handed $\Tilde{t}_1$, is given by~\cite{Kane:2019nes}:
\begin{eqnarray}
  \nonumber  \gamma_{CP,\Tilde{g}}&=&\frac{\Gamma(\Tilde{g}\rightarrow t+s+b)-\Gamma(\Tilde{g}\rightarrow \bar t+\bar s+\bar b)}{\Gamma(\Tilde{g}\rightarrow t+s+b)+\Gamma(\Tilde{g}\rightarrow \bar t+\bar s+\bar b)+\Gamma(\Tilde{g}\rightarrow \Tilde{B}+t+\bar t)}\\
    &\simeq& 10^{-4}\left(\frac{m_{\Tilde{g}}}{5\,\text{TeV}}\right)\left(\frac{m_{\Tilde{B}}}{1\,\text{TeV}}\right)\left(\frac{10\,\text{TeV}}{m_{\tilde{t}}}\right)^2 f\left(\frac{m^2_{\Tilde{B}}}{m^2_{\Tilde{g}}}\right)\sin(2\phi_{12}),
\end{eqnarray}
where kinematic factor $f(x)\equiv1-8x+8x^3-x^4-12x^2\,\text{ln}\, x$~\cite{Kane:2019nes} takes values between 0 and 1 and and $\phi_{12}$
is the phase difference between $\Tilde{g}$ and $\Tilde{B}$.
The total baryon asymmetry produced after the modulus decay through gluinos is given by:
 \begin{eqnarray}
   \nonumber Y_B&=&\frac{3}{4}\times 2\text{BR}(X\rightarrow \Tilde{g}\Tilde{g})\gamma_{CP,\Tilde{g}}\frac{g_{\star}(T_\text{rh})}{g_{\star,S}(T_\text{rh})}\frac{T_\text{rh}}{m_X}\\
    &\simeq&\frac{3\times 5^{1/4}}{2\pi^{3/4}}
     \frac{g_{\star}(T_\text{rh})^{3/4}}{g_{\star,S}(T_\text{rh})}\gamma_{CP,\Tilde{g}}\text{BR}(X\rightarrow \Tilde{g}\Tilde{g})
     (1-\beta)^{1/4}c^{1/2}
     \sqrt{\frac{m_X}{M_\text{Pl}}}.
 \end{eqnarray}
 Therefore, a modulus filed with a mass of $100\,\text{TeV}\lesssim m_X\lesssim10^{4}\,\text{TeV}$ which dominates the energy density of the Universe, can generate the baryon asymmetry of the Universe via decay into gluinos for large relative gaugino phase. The decay of the modulus field reheats the Universe to a temperature of $1\,\text{MeV}\lesssim T_\text{rh}\lesssim1\,\text{GeV}$  at which the washout effects are irrelevant. PBHs with a maximum mass of $10^{29}\,\text{g}\lesssim M_\text{PBH, max}\lesssim8\times 10^{35} \,\text{g}$ might form during the EMDE caused by the modulus field and they can explain DM relic abundance today for $3\times 10^{-10}\lesssim\beta\lesssim 3\times 10^{-7}$. To overcome the constraints on the mass of PBH as DM candidate, an extended mass function caused by power spectrum of primordial density fluctuations needs to be considered.
Heavier modular fields can also lead to the right amount of baryon asymmetry at higher reheating temperatures through gluino decay into quark and squark while the effects of washout needs to be considered~\cite{Ishiwata:2013waa}.

Since baryogenesis by modulus decay is induced through R-parity violating operators, the lightest superparticle is not stable and cannot be a DM candidate. 
This downside can be overcome by formation of heavy PBHs prior to the decay of the modulus field; a modular cosmology provides the proper environment for formation of heavy PBHs as DM candidate and can also explain baryon asymmetry successfully.

\section{CONCLUSIONS}
\label{sec:conclusion}
In this paper we have proposed a novel framework that utilises a heavy metastable particle as the common origin of baryon asymmetry in the Universe and DM which is composed of PBHs. 
The heavy metastable particle that comes to dominate the energy density of the Universe before its decay, gives rise to an EMDE. Within this era, formation of PBHs enhances and these PBHs might be heavy enough to survive to the present time and contribute to the abundance of DM.
The EMDE eventually transits to a radiation-dominated epoch at the lifetime of the heavy metastable particle. Out of thermal equilibrium decay of the heavy metastable particle, in the presence of baryon number and $CP$ violating interactions, can explain baryon asymmetry in the Universe. 
In this scenario, the ratio of the relic abundance of DM (PBH) to the relic abundance of baryon depends on the small initial abundance of PBHs and the large hierarchy between mass of the heavy metastable particle and mass of proton. These two factors can potentially counterbalance each other and make the final relic abundances of DM and baryons comparable.

We started with three free parameters $m_X, \tau_X, \beta$ and assume a reasonable choice of $CP$ violation e.g. $\gamma_{CP} = 10^{-2}.$ 
Fig.~\ref{fig:parameterspace} shows a variety of constraints we have imposed on these parameters.
In order for the PBHs to explain the full DM abundance of our Universe, the parameters must lie along the inclined right-hand edge of the red region. Then, for a given choice of any one of the three parameters $m_X, \tau_X, \beta$, the other two are also determined in order to explain the DM and baryogenesis of our Universe. Then we automatically (and not surprisingly) find the baryonic and DM contributions to the mass density of the Universe to be comparable.
As we showed, due to the constraints, the parameter space that works is restricted. Adding extra degrees of freedom such as a particle DM candidate can enlarge the allowed parameter space. 

As an example, we have considered an EMDE triggered by a modulus field. The decay chain of modulus into gluino, together with R-parity violating interactions, can explain the baryon asymmetry in the Universe. While the lightest superparticle cannot be a viable DM candidate, PBHs formed before decay of the modulus field can be heavy enough to contribute to the relic density of DM at the present time.    
 
\acknowledgments
We thank Katherine Freese for helpful discussions and comments on the manuscript. The work of B.S.E. is supported in part by DOE Grant DE-SC-0002424. 
\bibliography{draft}{}
\end{document}